\documentclass{emulateapj}
\usepackage{color}
\usepackage[normalem]{ulem}  

\newcommand{\noun}[1]{\textsc{#1}}
\newcommand{\dens}{g~cm$^{-3}$}

\slugcomment{Not to appear in Nonlearned J., 45.}

\shorttitle{Type Ia supernovae: Can Coriolis force break the symmetry of the GCD mechanism?}

\shortauthors{Garc\'\i a-Senz, Cabez\'on, Dom\'\i nguez \& Thielemann}

\begin{document}


\title{Type Ia supernovae: Can Coriolis force break the symmetry of the Gravitational Confined Detonation explosion mechanism?}  


\author{D. Garc\'\i a-Senz\altaffilmark{1,2},   
R.M. Cabez\'on\altaffilmark{3},
I. Dom\'\i nguez\altaffilmark{4},
F.~K. Thielemann\altaffilmark{3}
}

\altaffiltext{1}{Departament de F\'\i sica i Enginyeria Nuclear, UPC, 
Compte d'Urgell 187, 08036 Barcelona, Spain; domingo.garcia@upc.edu} 
\altaffiltext{2}{Institut d'Estudis Espacials de Catalunya, Gran Capit\`a 2-4, 
 08034 Barcelona, Spain}
\altaffiltext{3}{Departement Physik, Universit\"at Basel. Klingelbergstrasse, 82, 4056 Basel, Switzerland; ruben.cabezon@unibas.ch}
\altaffiltext{4}{Departamento de F\'\i sica, Te\'orica y del Cosmos, Universidad de Granada, 18071 Granada, Spain}  



\begin{abstract}

Nowadays the number of models aimed at explaining the Type Ia supernova phenomenon is high and discriminating between them is a must-do. In this work we explore the influence of rotation in the evolution of the nuclear flame which drives the explosion in the so called gravitational confined detonation models. Assuming that the flame starts in a point-like region slightly above the center of the white dwarf (WD) and adding a moderate amount of angular velocity to the star we follow the evolution of the deflagration using a smoothed particle hydrodynamics code. We find that the results are very dependent on the angle between the rotational axis and the line connecting the initial bubble of burned material with the center of the white dwarf at the moment of the ignition. The impact of rotation is larger for angles close to $90^0$~because the Coriolis force on a floating element of fluid is maximum, and its principal effect is to break the symmetry of the deflagration. Such symmetry breaking weakens the convergence of the nuclear flame at the antipodes of the initial ignition volume, changing the environmental conditions around the convergence region with respect to non-rotating models. These changes seem to disfavor the emergence of a detonation in the compressed volume at the antipodes, thus compromising the viability of the so called gravitational confined detonation mechanism.

\end{abstract}




\section{Introduction}

Because of the many connections of Type Ia supernovae (SNe Ia) with fundamental problems in Astrophysics such as, for example, the chemical evolution of galaxies \citep{iwa99} or the origin of the acceleration of the universe \citep{rie98}, the quest of the mechanism behind the explosion is of utmost relevance. There is a general agreement that the supernova display involves the explosion of a white dwarf but beyond this point there is not an agreement on the details. The explosion may be the outcome of the destabilization of a white dwarf which approaches the Chandrasekhar-mass limit owing to the mass accretion from a nearby companion star, a succession of events referred as the  single degenerated scenario (SD) \citep{whe73,nom82}. Another route to the explosion involves the merging of two white dwarfs with more or less canonical masses $M_{WD}\simeq 0.7$~M$_\sun$~orbiting in a close binary system \citep{ibe84,lor09,pak13}, known as the double degenerate scenario (DD), which includes the direct collision of two WD as a limiting case \citep{ros09a,gar13}. Actually, both channels may coexist and contribute to the observed averaged rate of SNe Ia \citep{wan12}. However, each scenario, SD or DD, exhibits a considerable degree of degeneracy because the basic observational properties of SNe Ia are  matched by different explosion models.
Excluding violent mergers and He-detonations, there is not a consensus about the mechanism driving the explosion \citep{bra95,hil00,hil13}. Among the different mechanisms found in the literature there are either pure (subsonic) deflagrations \citep{nom84,fin14} or deflagrations which at some critical density turn into a (supersonic) detonation \citep{kho91a}, although the physical agent driving the deflagration to detonation transition is uncertain. Another way to detonate the WD relies on the gravitational confinement of the plasma (the cosmic version of the terrestrial inertial confinement fusion). In these confinement scenarios  there is a brief deflagrative phase which causes the expansion of the WD before a large amount of material is burnt. The explosion is rather weak which results in a failed SNe Ia. Still the explosion can be revitalized in two ways leading to the pulsating and gravitational confined models of thermonuclear supernova respectively. The pulsating explosion models arise after the fallback of some material previously expelled and the formation of an accretion shock which compresses and heats the core. In steady conditions the ram pressure exerted by the infalling material keeps the plasma compressed long enough to keep ignited the $^{12}C+^{12}C$~reaction, which under the appropriate physical conditions, may give rise to a Chapman-Jouguet detonation which incinerates the whole core. Because in these scenarios the explosion ensues after a  global pulsation of the WD, they were called Pulsating Delayed Detonation models (PDD) in spherically symmetric calculations \citep{iva74,kho91b} and Pulsating Reverse Detonations (PRD) in three dimensions \citep{bra06,bra09,brb09}. A second possibility is that considered in the gravitational confined detonation models (GCD) \citep{ple04,ple07,rop07,tow07,jor08,mea09,sei09} whose main features are: 1) the deflagration flame ignites in a small volume offset from the center of the WD, 2) the bubble of hot fluid accelerates vertically by buoyancy in the strong gravitational field of the compact object, 3) soon the bubble made of ashes expands over the surface of the star and converges at the opposite side from where the initial breakout occurred, 4) as the convergence at the antipodes is strong it produces an inwardly moving jet which may give rise to the detonation in (or close to) the core.

Until now all GCD models have been calculated assuming that the white dwarf is not rotating at the moment of the explosion. Nevertheless, we know that in both, the SD and DD scenarios, the WD could store an important amount of rotation because of the angular momentum transferred from the accretion disc to the WD \citep{pie03,yoo04,yoo05}. In GCD models the deflagrative phase sets the conditions of the detonation at late times, and it is very likely that the rotation axis is not perfectly aligned with the symmetry plane of the flame. As a consequence, Coriolis force will act with different strengths in different regions of the burning material. Therefore, to what extent may rotation weaken the efficiency of the convergence of the ashes at the antipodes?

Although common proposed scenarios of SNe Ia imply the rotation of a WD, the number of  multidimensional calculations of the explosion which incorporates rotation is really scarce. One of them was carried out by \citet{ste92}, simulating the detonation of a fast rotating WD. More recently, \citet{pfa10a} have calculated the deflagration of a rapidly spinning WD in three dimensions, after the central ignition of the fuel, and conclude that the yields of the species synthesized during the explosion do not match the observed spectra. In a subsequent work the same authors \citep{pfa10b} analyzed the  detonation of a fast rotating WD and conclude that this mechanism could explain the existence of some superluminous SNe Ia.

In the present work we analyze for the first time the possible consequences of including the rotation in the GCD scenario. To do that we have carried out several 3D simulations using a smoothed particle hydrodynamics (SPH) code with moderate resolution. The amount of rotation considered in our exploratory study is modest $\Omega/\Omega_c\simeq 0.08$, where $\Omega_c\simeq 4.7$~s$^{-1}$~is the critical Keplerian velocity at the equator, to not appreciably modify the spherically symmetric density profile of the WD at the moment of the explosion. As we will see, even that small amount of rotation is able to change the thermodynamic conditions at the convergence point of the ashes. Therefore, future studies dealing with confinement scenarios should incorporate rotation as a basic ingredient,  adding a step of difficulty to an already complex problem. 

In Section 2 we describe the main features of the hydrocode, the initial setting and the method to calculate the evolution of the nuclear flame (described with more detail in the Appendix). We also provide here a brief discussion of the profile of the angular velocity adopted in the simulations. A detailed description of the hydrodynamic evolution of the four considered models is given in Section 3. Finally, Section 4 summarizes the main conclusions of our work.

\section{Hydrodynamic method. Flame handling and initial setting} 

In the GCD and PRD scenarios, the final explosion of the white dwarf takes place after a considerable change in the WD radius. Being a Lagrangian method free of numerical diffusion, SPH is well suited to handle systems hosting a variety of dynamical scales \citep{ros09b,spr10} and to track the complex geometry of the nuclear flame which powers the explosion. As angular momentum is exactly conserved in SPH, this technique is also adequate to handle problems involving rotation. We are using an updated version of the SPH code devised more than one decade ago which was applied to simulate the thermonuclear explosion of a WD by different physical mechanisms either deflagrations \citep{gar05} or detonations \citep{bra08}. Nevertheless, our new hydrocode \noun{SPHYNX}~(Cabez\'on \& Garc\'\i a-Senz, 2015 in  preparation), incorporates a large number of state-of-the-art improvements that are worth to mention. The most relevant update concerns gradient estimation, which now relies on an integral approach which is more accurate than the traditional method in SPH \citep{gar12,cab12}. To carry out interpolations \noun{SPHYNX} makes use of the $sinc$~family of kernels \citep{cab08}, which are more resistant to particle clustering than the standard cubic spline, therefore allowing to increase the number of interpolating particles in the SPH summations to reduce the numerical noise. The smoothing length $h({\bf r},t)$~is updated according to the method by \cite{spr03}, which ensures both, energy and entropy conservation. In addition the code has been parallelized with an hybrid scheme in MPI + openMP, so that we are routinely using $\simeq 2\times 10^6$~particles.   

The algorithm to track the flame still relies on a diffusion-reaction scheme but it incorporates some novel improvements. Among them, a phenomenological subgrid model to estimate the characteristic turn-over velocity, $v_{t}$, of the eddies at scales resolved by the hydrocode (Appendix A). The effective velocity of the flame is $v_{f}=max(v_l, v_{t})$~$v_l$, being the laminar velocity of the flame  from \cite{tim92}. At the typical conditions considered in our simulations we obtain $100$~km~s$^{-1}\lesssim v_{f}\lesssim 500$~km~s$^{-1}$, the former being roughly the conductive flame velocity at $\rho\simeq 2\times 10^9$~\dens~and the  latter only achieved in points where the shear is high. Different groups approximate $v_{f}$~using a variety of numerical schemes, based either on a careful estimation of the turbulent velocity plus the level-set approach \citep{rei99}, or in diffusion-reaction schemes either with constant propagation velocity \citep{gar05,maa13} or powered by the Rayleigh-Taylor instability \citep{gam99,ple04,ple07}. In any case, the gross details of the evolution are insensitive to the precise value of $v_{f}$~because the effective flame velocity self-regulates through the creation/destruction of the heat-exchanging surface between ashes and fuel \citep{kho95}.

The physics included are very similar to those recently used by \cite{gar13} to simulate the collision of white dwarfs. The nuclear network is an $\alpha-$chain complemented with carbon and oxygen binary reactions. The evolution of the species is calculated implicitly with the temperature to ensure a smooth transition to the nuclear-statistical equilibrium (NSE) regime \citep{cab04}. Electron captures on protons and nuclei have been neglected because their impact on the dynamics of the explosion is secondary \citep{mea09}. Our EOS has the contributions of electrons \citep{bli96}, ions (including Coulomb and polarization corrections) and radiation.                               

The initial model is a white dwarf with central density $\rho_c=2.6\times 10^9$~\dens, homogeneous temperature $T=2\times 10^7$~K and composition $X_C=X_O=0.5$. The integration of the Lane-Emden equation for these conditions gives a mass $M_{WD}=1.37$~M$_{\sun}$. The WD was then mapped to a three-dimensional configuration of particles and relaxed to get rid of the excess of numerical noise before starting the hydrodynamic simulations. The maximum resolution is $\simeq 12$~km, achieved at the center of the WD. While this resolution is high enough to capture the main features of the explosion during the deflagration phase it is below of what is needed to study the transition to a possible detonation at late times.

All calculations reported in this paper assume that the thermonuclear ignition of the WD starts in a single small region (a bubble) close to the center of the star and propagates to the remaining plasma by hydrodynamic instabilities. This initial setting is similar to that currently used by other groups easing the comparison of results \citep{rop07,mea09}. On another note, the ignition in a single region at an altitude $\simeq 50$~km is favored by the recent studies by \citet{zin11,non12,mal14}~of the preignition stage using the MAESTRO code. In several of these works slow rotation was shown to affect the long convective phase preceding the ignition \citep{zin11}, (see also \citealt{kuh06}). We also assume that before the convergence of the ashes the combustion regime is purely deflagrative, that is we have ignored any potential deflagration
 to detonation transition when the flame enters into the distributed regime at densities $\rho\simeq 2\times 10^7$~\dens \citep{kho91a,gam05}.

The main goal of this work is to explore the impact of rotation on the GCD models. Nevertheless, we have restricted the study to moderate rotators to preserve the spherical symmetry and equilibrium properties of the initial model, which is the same for all calculated models. Considering faster rotators would need a totally different initial setting, being beyond the scope of the present manuscript. The main features of the four calculated models during the deflagration phase are given in Table~\ref{table1}.   

The rotational velocity of a WD at the end of the accretion phase is poorly known.
There is some agreement on that the  core is probably rotating as a rigid body  and, beyond that, the angular velocity may increase up to a critical radius $R_c$, to finally decay to Keplerian values at the surface of the WD \citep{yoo04,pfa10a}. Nevertheless in those cases where an efficient redistribution of angular momentum is assumed the WD may approach rigid rotation \citep{pie03}. Rigid rotators could be the result of having either strong braking magnetic torques or efficient viscous angular momentum transport and/or long time-scales \citep{pir08}.  

 In our exploratory study we have considered four models (Table~ \ref{table1}): no rotation, rigid rotators with $\Omega_z=0.4$~s$^{-1}$~and $\Omega_x=0.4$~s$^{-1}$~respectively, and $\Omega_x$, 

\begin{equation}
\Omega_x=
\left\{\begin{array}{rclcc}
\Omega_0 = 0.6 & \quad r \le R_c,\\
\frac{\Omega_0}{R_c-R_{WD}}\left(r-R_{WD}\right) & \qquad r > R_c, 
\end{array}
\right.
\label{omega_1}
\end{equation}
    
\noindent
where $r$~is the distance to the center and $R_c=5\times 10^7$~cm. This last case is a simple profile consisting of a rigid body rotation near the center which decays linearly to a much lower value at the surface. The case $\Omega_z\ne 0$~sets the rotating axis aligned to the initial ascending direction of the bubble, whereas $\Omega_x\ne 0$~takes that axis orthogonal to the initial displacement of the bubble. It is in the latter case where we expect the maximum effect of the rotation, mainly driven by the Coriolis force. The value of this acceleration, ${\bf a}_{Cor}=2{\bf v}\times {\bf \Omega}$~acting on a piece of a floating bubble is one of the greatest Coriolis forces found in Nature. To have an estimate of the order of magnitude, we can take $\Omega=0.4$~s$^{-1}$ (models B and C of Table~\ref{table1}) and a typical rising velocity of a bubble $v\simeq 2.5\times 10^8$~cm~s$^{-1}$. This yields ${\bf a}_{Cor}\simeq 2\times 10^8$~cm~s$^{-2}$. A huge value, almost comparable to the effective acceleration of the bubble imparted by the Archimedean force $a_{eff}= g\frac{\Delta\rho}{\rho}$.

\begin{figure}
\includegraphics[angle=-90,width=\columnwidth]{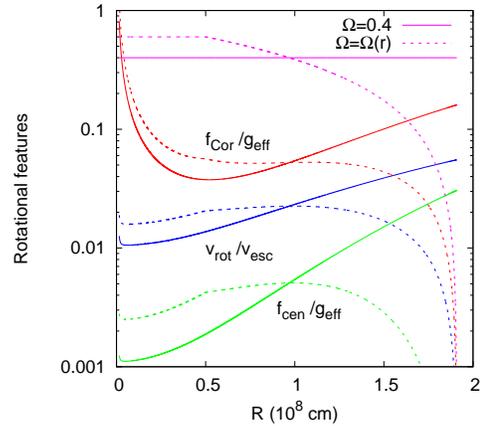}
\caption{Dimensionless indicators of the equatorial rotational strength for two choices of the angular velocity $\Omega$, of the WD. Continuum lines are for rigid rotators with $\Omega=0.4~s^{-1}$. Dashed lines correspond to $\Omega(r)$~calculated with Equation~\ref{omega_1}.  
}
\label{fig1}
\end{figure}

\begin{figure*}
\includegraphics[angle=270,width=\textwidth]{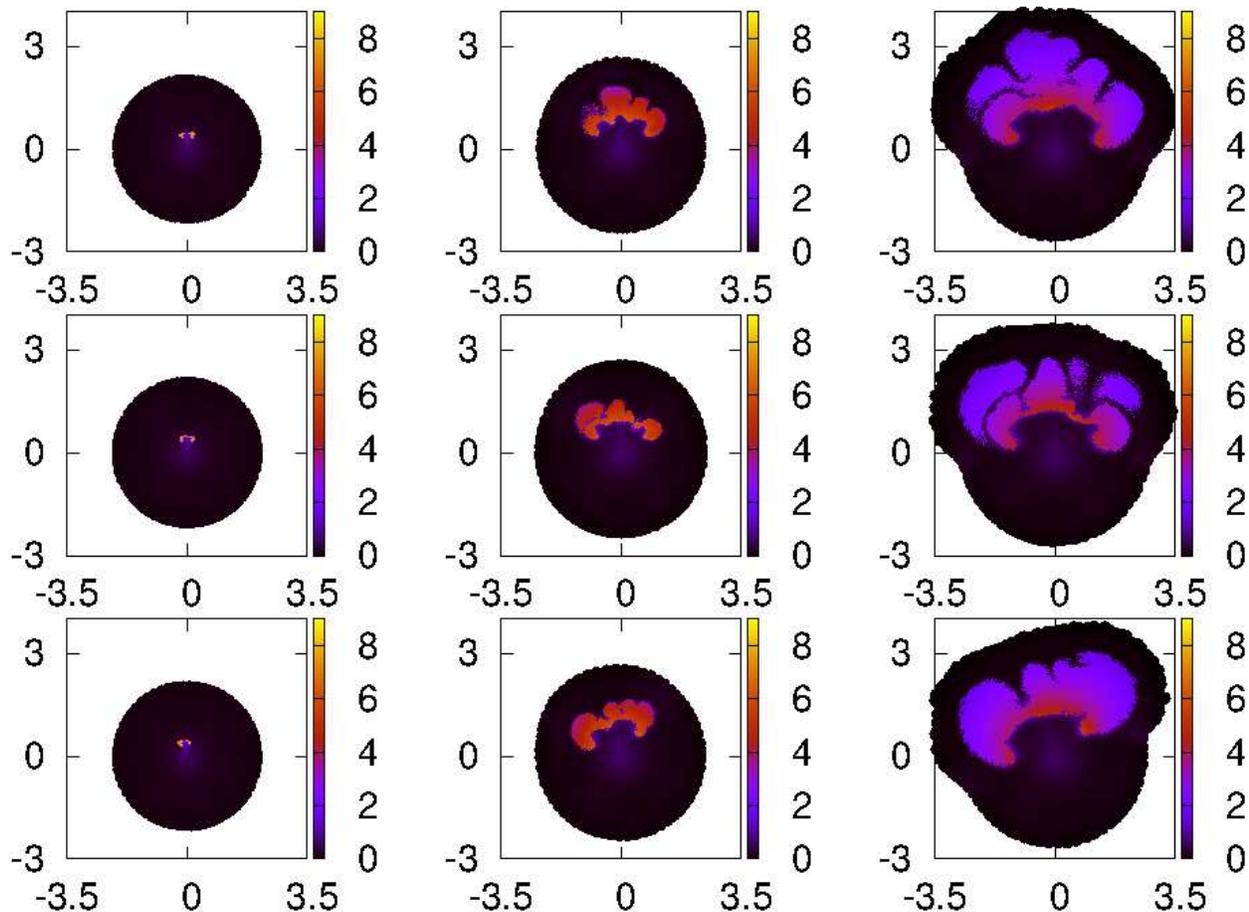}
\caption{YZ-slice showing a color-map of temperature (in units of $10^9$~k) for models A (upper row), B (central row), C (lower row) in Table~\ref{table1} at times $t= 0.59$~s, $t=0.97$~s and $t=1.18$~s respectively. Coordinates are labeled in units of $10^8$~cm.  
}
\label{fig2}
\end{figure*}

\begin{figure}
\includegraphics[angle=-90,width=\columnwidth]{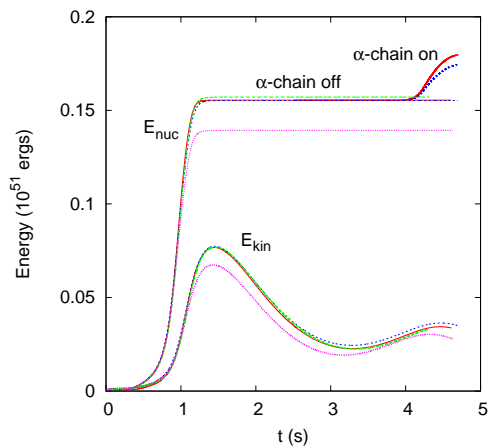}
\caption{Evolution of the kinetic energy and the released nuclear energy for models A (continuum red line), B (long-dashed line in green), C (dashed-blue line), and D (dots in pink). After $t=2.5$~s nuclear reactions were either switched-off or kept switched-on (thick lines). 
}
\label{fig3}
\end{figure}

\begin{figure*}
\includegraphics[angle=270,width=\textwidth]{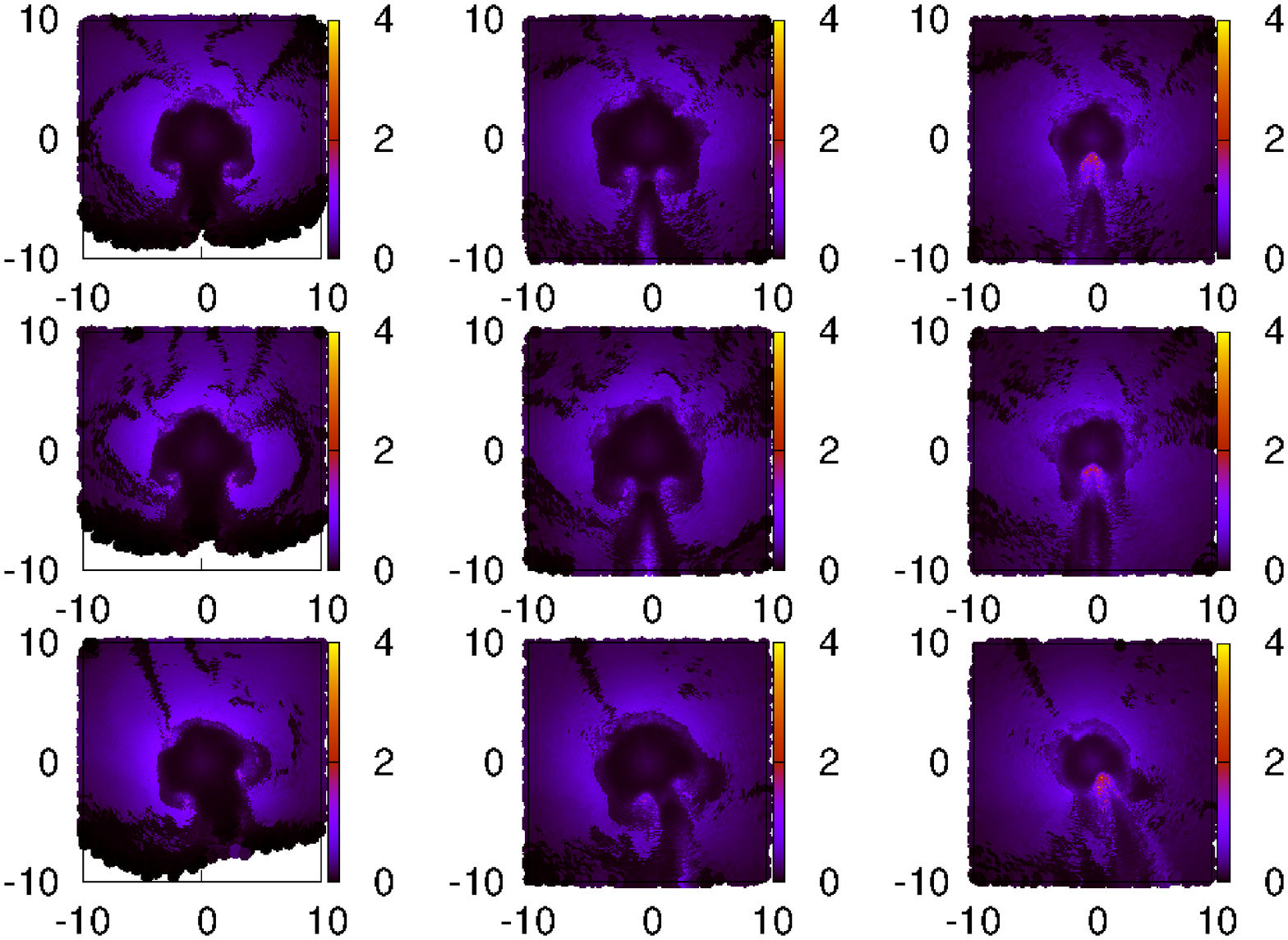}
\caption{Same than Figure~\ref{fig2} but at times $t= 2.5, 3.43, 4.25$~s respectively. Because the WD has growth in size only the innermost part of the object is depicted to highlight the convergence of the ashes towards the antipodes. 
}
\label{fig4}
\end{figure*}

In  Figure~\ref{fig1} we show several magnitudes related to rotation. As we can see, neither the rotational linear velocity nor the centripetal force are large enough to significantly break the spherical symmetry of the initial model. The largest ratio between the linear velocity and the local escape velocity is $\simeq 5\%$, obtained just on the equator at the surface of the WD. The ratio between the centripetal force and the floating force acting on a rising blob $g_{eff}=(\Delta\rho/\rho) g(r)\simeq 0.2~g(r)$, always remains below $3\%$. Thus, the perturbation introduced by a moderate amount of rotation hardly affects the stability of the initial model during the first tenths of second of flame propagation. At longer times the nuclear energy released during the explosion outpowers the kinetic rotational energy by several orders of magnitude. The ratio between the rotational kinetic energy and the binding energy of the WD at $t=0$~is also small, $E_{rot}/E_{bin}\le 0.2\%$~in all models. However, the ratio between the Coriolis and floating forces can be as high as $10\%$~during the first tenths of second of flame propagation and should not be disregarded. Moreover, the Coriolis force is pushing the bubble all the time during its race to the antipodes and the cumulative effect may break the symmetry of the convergence of the ashes, especially when the rotational axis is not aligned with the initial line of displacement of the bubble.      

\begin{deluxetable*}{lrrrrcrrrrrrr}
\tablewidth{40pc}
\tablecaption{Main features of both the initial models and the deflagration phase \label{table1}}
\tablehead{
\colhead{Model}& \colhead{$z_0$}&\colhead{$r_b^0$}& \colhead{$\Omega_x$} & \colhead{$\Omega_y$}&\colhead{$\Omega_z$}&\colhead{E$_{nuc}$}&\colhead{M$_{Si}$}&\colhead{M$_{Ni}$}& \colhead{$\rho_c(min)$}&\colhead{$h_c$}&\colhead{$\rho_{jet}(T_9=3)$}&\colhead{$t(T_9=3)$} \cr
 & \colhead{km}& \colhead{km}&\colhead{$s^{-1}$}&\colhead{$s^{-1}$}&\colhead{$s^{-1}$}&\colhead{$10^{51}$~ergs}&\colhead{$M_\sun$}&\colhead{$M_\sun$}&\colhead{$10^7$\dens}&\colhead{km}&\colhead{$10^7$~\dens}& \colhead{s} 
}
\startdata
A&$60$&$40$ &$0.0$&$0.0$& $0.0$&$0.155$&$2.5\times 10^{-3}$&$9.0\times 10^{-2}$&$4.6$&$49$&$1.78$&$4.24$ \\
B&$60$&$40$ &$0.0$&$0.0$& $0.4$&$0.157$&$2.6\times 10^{-3}$&$9.1\times 10^{-2}$&$4.5$&$49$&$1.36$&$4.28$ \\
C&$60$&$40$ &$0.4$&$0.0$& $0.0$&$0.155$&$3.1\times 10^{-3}$&$8.9\times 10^{-2}$&$4.7$&$48$&$1.16$&$4.18$ \\
D&$60$&$40$ &$\le 0.6$&$0.0$& $0.0$&$0.139$&$2.3\times 10^{-3}$&$8.1\times 10^{-2}$&$5.7$&$41$&$1.80$&$4.01$ \\
\enddata
\tablecomments{The meaning of of the columns is as follows: $z_0$~ignition altitude; $r_b^0$~initial radius of the bubble; $\Omega_x, \Omega_y, \Omega_z$~components of the angular velocity at t=0 s; $E_{nuc}$~released nuclear energy; $M_{Si}, M_{Ni}$~yields of silicon and nickel; $\rho_c(min), h_c$~minimum density and resolution achieved at the center of the core of the WD; $\rho_{jet}(T_9=3), t(T_9=3)$~density in the colliding region when the temperature of a particle made of fuel exceeds $3\times 10^9$~K and elapsed time to reach that temperature.
}
\end{deluxetable*}

\section{Hydrodynamic simulations}

\subsection{The deflagration phase}

\begin{figure}
\includegraphics[angle=-90,width=\columnwidth]{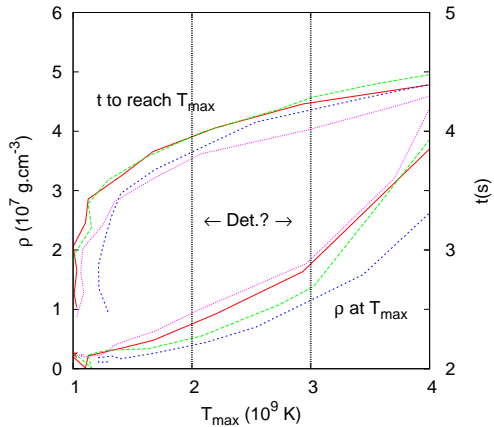}
\caption{Density of the particle as a function of the instantaneous maximum value of temperature of fuel particles $T_{max}$~achieved in the collision region (with nuclear reactions turned off). Model A (continuum line in red), model B (dashed line in green), model C (short-dashed line in blue) and model D (dots in pink). The elapsed time needed to reach $T_{max}$~(uppermost curves) can be read on the vertical axis at the right. The vertical bars in black shows the region where the transition to a detonation may occur.  
}
\label{fig5}
\end{figure}

\begin{figure*}
\includegraphics[angle=270,width=\textwidth]{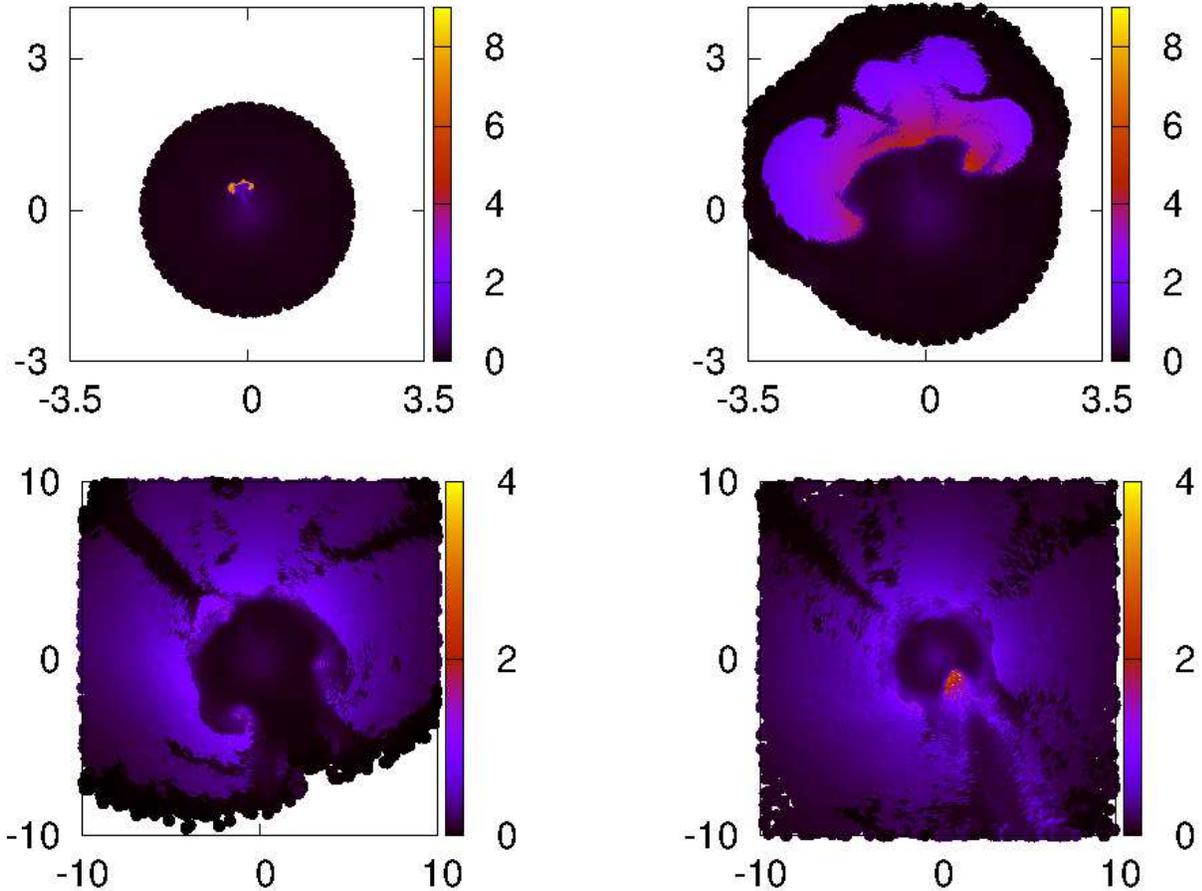}
\caption{Same as Figures~\ref{fig2} and \ref{fig4} but for model D at times 
 $t= 0.6, 1.21$~s  (upper rows) and $t=2.54, 4.21$~s (lower rows). 
}
\label{fig6}
\end{figure*}

\begin{figure*}
\includegraphics[angle=270,width=\textwidth]{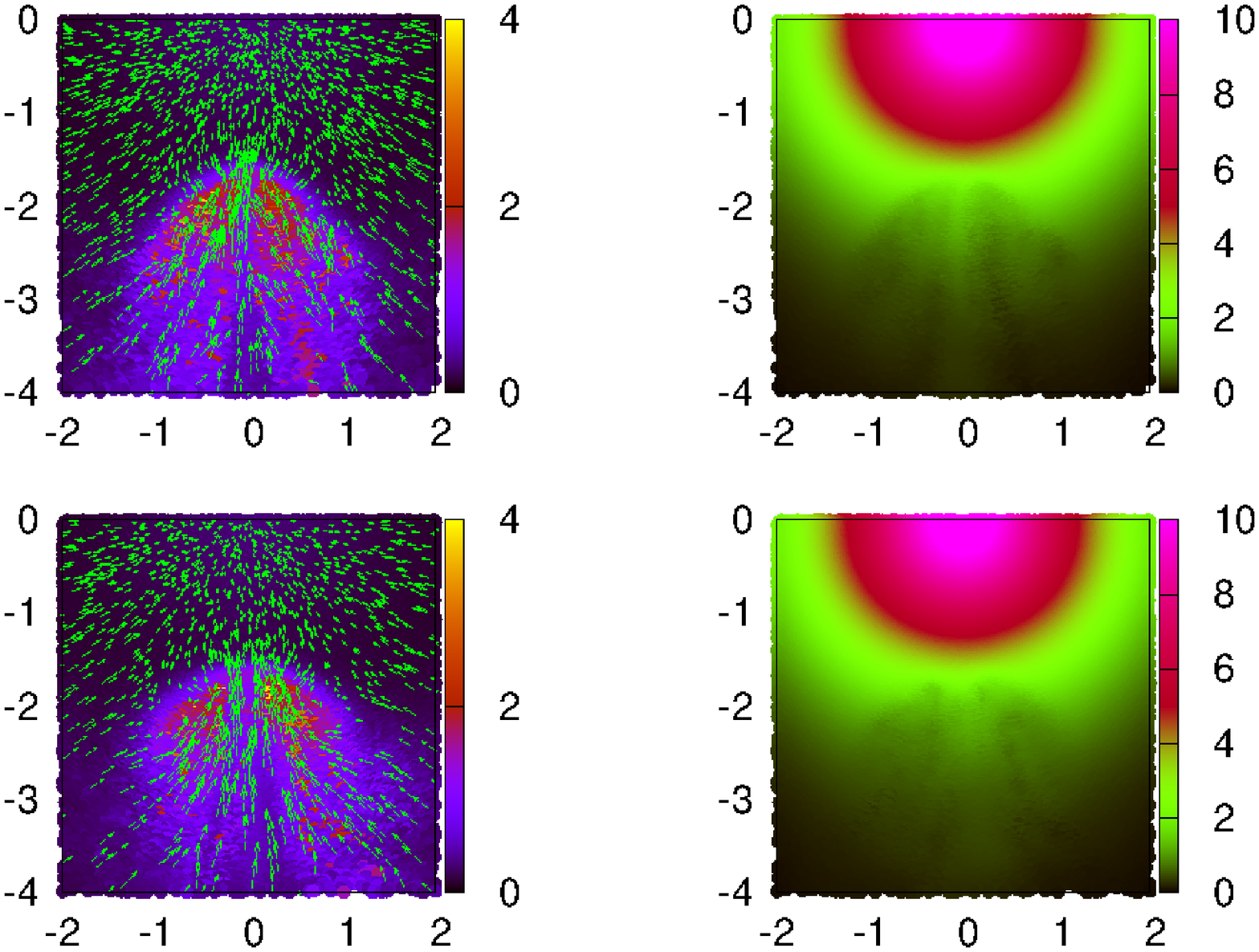}
\caption{Detail of the collision region for model A at time $t=4.229$~s (upper panels) and model B at $t=4.289$~s (lower panels). Left column shows the velocity field superimposed to the temperature (in units of $10^9$~K) color-map. Right column depicts density in units of $10^7$~\dens.
}
\label{fig7}
\end{figure*}

Model A in Table~\ref{table1} was calculated assuming that the WD is not rotating when the first sparks ignite, and serves as a reference model. It is also assumed that these seminal sparks ignite in a point-like region close to (but not at) the center of the core. In all simulated models the geometry of the ignited region is almost spherical with an initial radius of 20 km. The center of mass of the hot bubble is located 60 km above the center of the WD. To provoke the ignition  the content of the bubble is isochorically burned to nuclear-statistical equilibrium (NSE), so that the temperature jumps to $\simeq 9$~billion degrees, giving birth to a thermal wave.  We let the flame grow  during $t=0.25~s$ keeping the WD structure frozen to ensure smooth conditions around the flame. At that time the radius of the hot volume has grown up to $40$~km. Afterwards, the time is reset to $t=0~s$~and the hydrodynamic calculation of the explosion starts. The bubble expands trying to regain equilibrium and its density drops below that of the surrounding plasma. It begins to float, acquiring a buoyant velocity $v\simeq 2600$~km/s at $t=0.5$~s. Soon the interaction of the rising blob with the environment moves its spherical geometry to a toroidal one (see the leftmost panels in Figure~\ref{fig2}). From this moment on the flame accelerates due to both, the increase of the effective surface between ashes and fuel, and to the strong vorticity induced by the Kelvin-Helmholtz instability at the edges of the rising blob. This last effect is handled using the toy model described in Appendix A. 

In Figure~\ref{fig3} we show the released nuclear energy, $E_{nuc}(t)$ and kinetic energy for all four calculated models. The evolution of models A, B and C is very similar whereas model D produces a bit less nuclear energy. The nuclear energy release shows a strong rise at $t> 0.8$~s and begins to saturate at $t\simeq 1.1$~s, owing to the expansion of the WD. At the elapsed time $t=1.5$~s all nuclear reactions are virtually quenched while the ashes are surfing around the core of the WD in their journey to the meeting point at the south pole, Figures~\ref{fig2}, \ref{fig4}. At $t=2.5$~s the nuclear reactions are {\sl switched-off} from the hydrocode to make the analysis of the convergence of the ashes easier. Although the injected energy $E_{nuc}\simeq 1.5\times 10^{50}$~ergs is not enough to unbind the star, it provokes a large change in its radius, and the central density drops to $\rho_c= 4.6\times 10^7$~\dens~at $t=3.02$~s (model A). The value of $E_{nuc}$~is in the range of values obtained in other 3D calculations with similar initial conditions \citep{rop07,fin14}. 

The stages that precede the collision of the ashes for models A, B and C are shown in Figure~\ref{fig4}. For the reference calculation without rotation (model A), the convergence of ashes takes place around the second snapshot at time $t= 3.43$~s. As we can see, despite the imprint of hydrodynamical instabilities there is an astonishing symmetry around the line joining the center of the remnant with the meeting point. The collision of the conical high-velocity stream of gas promotes the birth of two jets moving to opposite directions (see the central panels in Figure~\ref{fig4}). The basic physics of jets born from the collision of fluid streams in both astrophysics \citep{ten88} and terrestrial laboratories \citep{har66} is well known. Depending on the angle of the collision, it may result in one (for low incident angles) or two jets moving in opposite directions. The latter is just what was reported in the simulations by \cite{ple07}, \cite{mea09} and can also be seen in the second and third columns of Figure~\ref{fig4}.  

According to Table~\ref{table1} the amount of $^{56}Ni$~synthesized during the deflagration is around $0.09$~M$_\sun$~with released nuclear energies $\simeq 0.15\times 10^{51}$~ergs. Thus, in those models where the igniting sparks are localized in a tiny, point-like region, a transition to a detonation (by the  GCD or the PRD mechanisms)  is necessary to raise the $^{56}Ni$~yields to standard observed values, $\simeq 0.6$~M$_\sun$. At this point it is worth noting that the analysis of the radioactive isotopes present in the debris of type Ia supernova SN2014J suggests the existence of a small amount of $^{56}Ni$~in the external layers of the exploding star \citep{die15}. Such early ejection could be spread in a truncated conical ring made of  $\simeq 0.05$~M$_\sun$~of $^{56}Ni$, detached from the main body of the ejecta (Isern et al. 2015, submitted). That amount of $^{56}Ni$~is curiously close to those shown in Table~\ref{table1}, which were produced during the deflagration of the WD. Although in our models the ejected $^{56}Ni$~is not exactly inside a truncated cone its distribution is biased towards the  hemisphere where the ignition of the WD starts.          

The evolution of the rotating models B, C and D primarily depends on the angle between the rotational axis and the initial line of flotation of the bubble. When that angle is zero the impact in the dynamics is small because 1) during the first tenths of second the bubble rises vertically, therefore the angle between $v_{b}$~and $\Omega$~is small and so is the Coriolis acceleration, and 2) $a_{Col}$~remains axisymmetric during the lateral expansion of the ashes preserving the symmetry of the flux. Thus, the evolution of model B is similar to model A, as it can be seen in the central row of panels in Figures~\ref{fig2} and \ref{fig4}. There are minor differences between models A and B, as for example the lower value of density achieved in model B at the head of the jet when a fiducial temperature, f.e. $T=3\times 10^9$~K, is crossed (column $12$ in Table~\ref{table1}). Also, the elapsed time needed to attain for the first time that temperature is larger for model B (column $13$ in Table~\ref{table1}).  

\begin{figure*}
\includegraphics[angle=270,width=\textwidth]{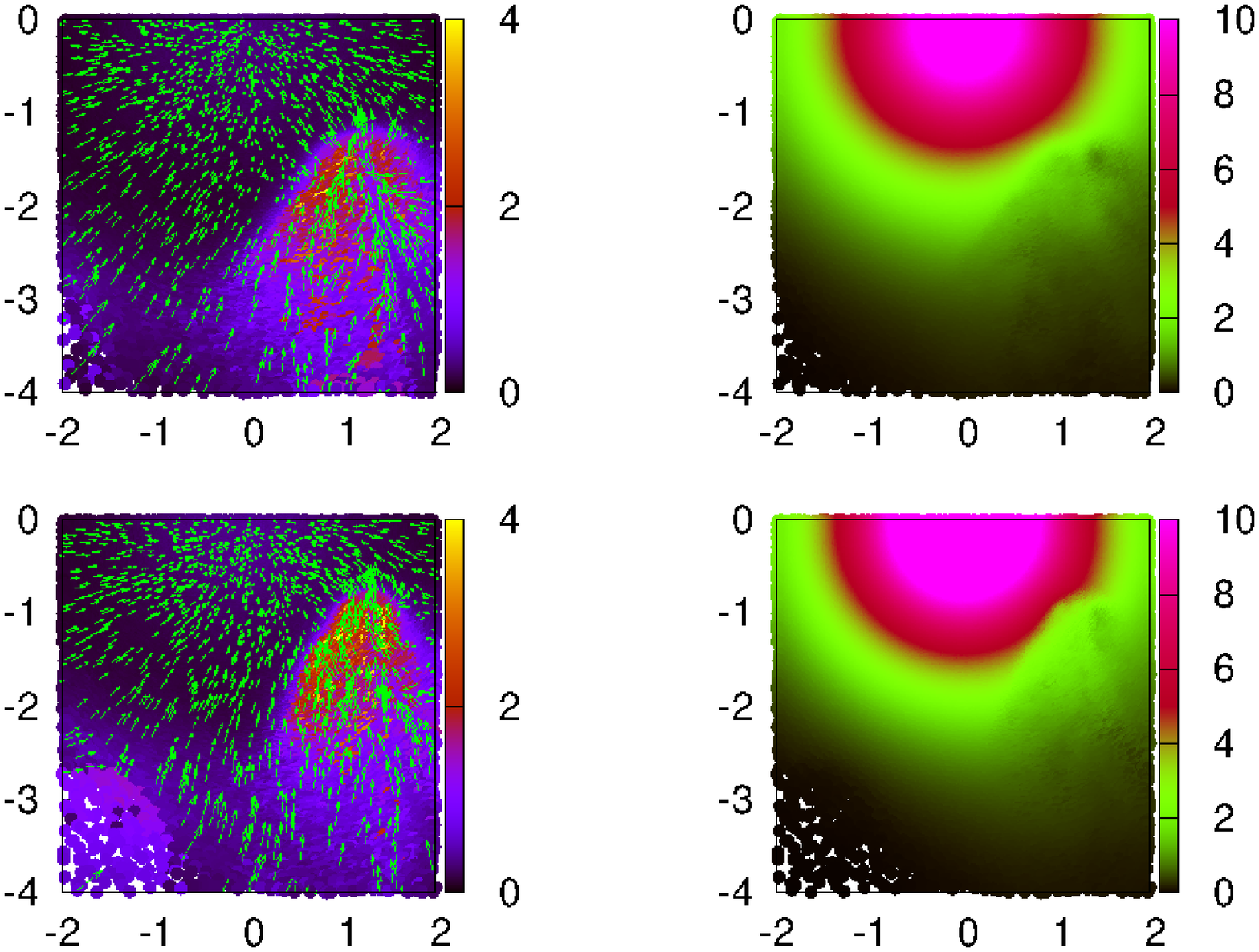}
\caption{Same as Figure~\ref{fig7} but for models C and D at times $t=4.277$~s and $t=4.21$~s. 
}
\label{fig8}
\end{figure*}

\begin{figure*}
\includegraphics[angle=-90,width=\textwidth]{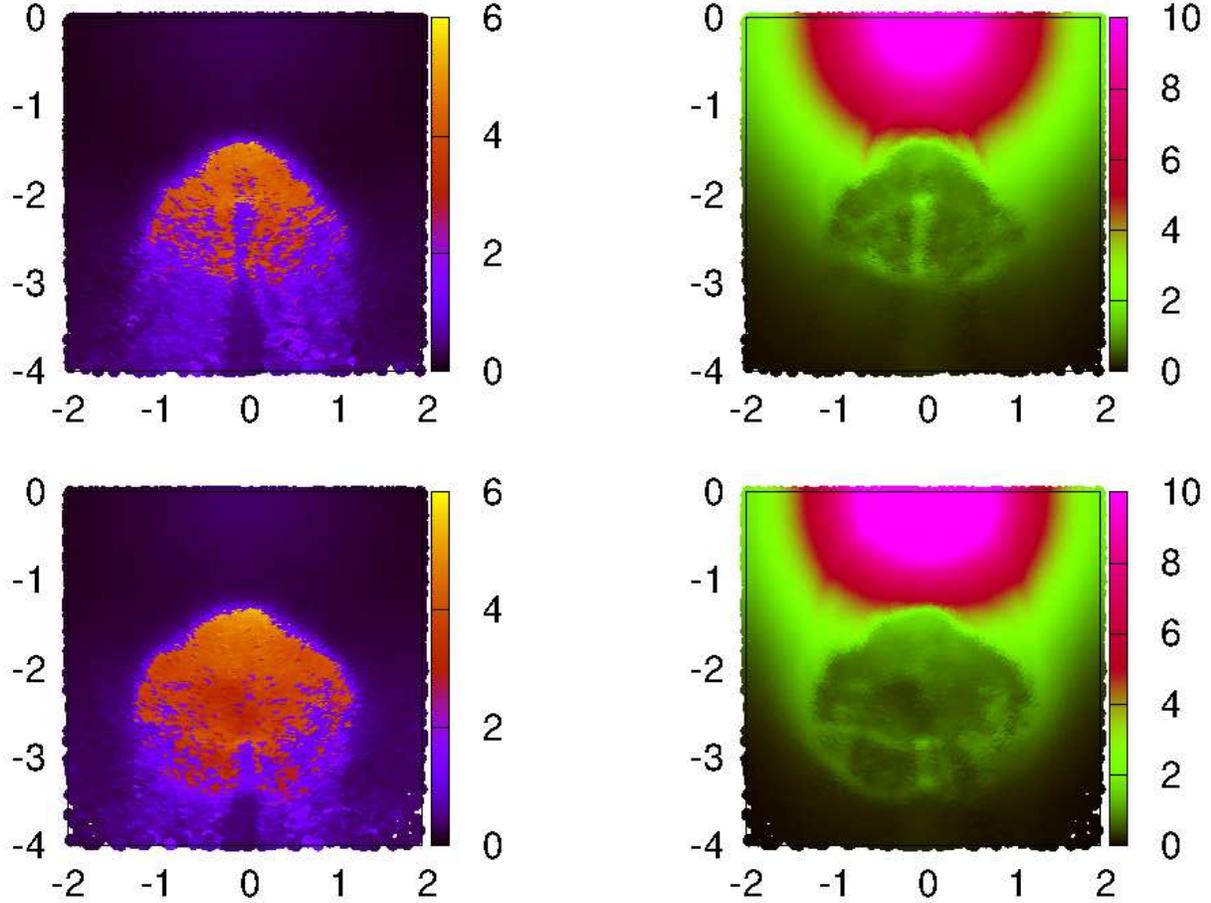}
\caption{Same as Figure~\ref{fig7} but for model A at times $t=4.28032$~s and $t=4.41304$~s with the nuclear reactions turned-on.  
}
\label{fig9}
\end{figure*}

\begin{figure}
\includegraphics[angle=-90,width=\columnwidth]{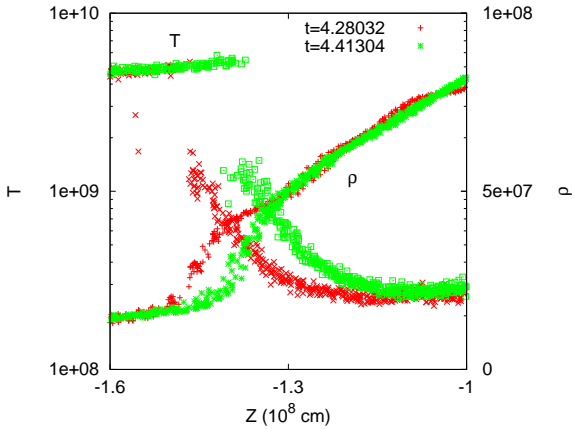}
\caption{Profiles of temperature and density of model A obtained from a one-dimensional cut around the symmetry line of the jet at times $t=4.28032$~s and $t=4.41304$~s.   
}
\label{fig10}
\end{figure}

\begin{figure*}
\includegraphics[angle=-90,width=\textwidth]{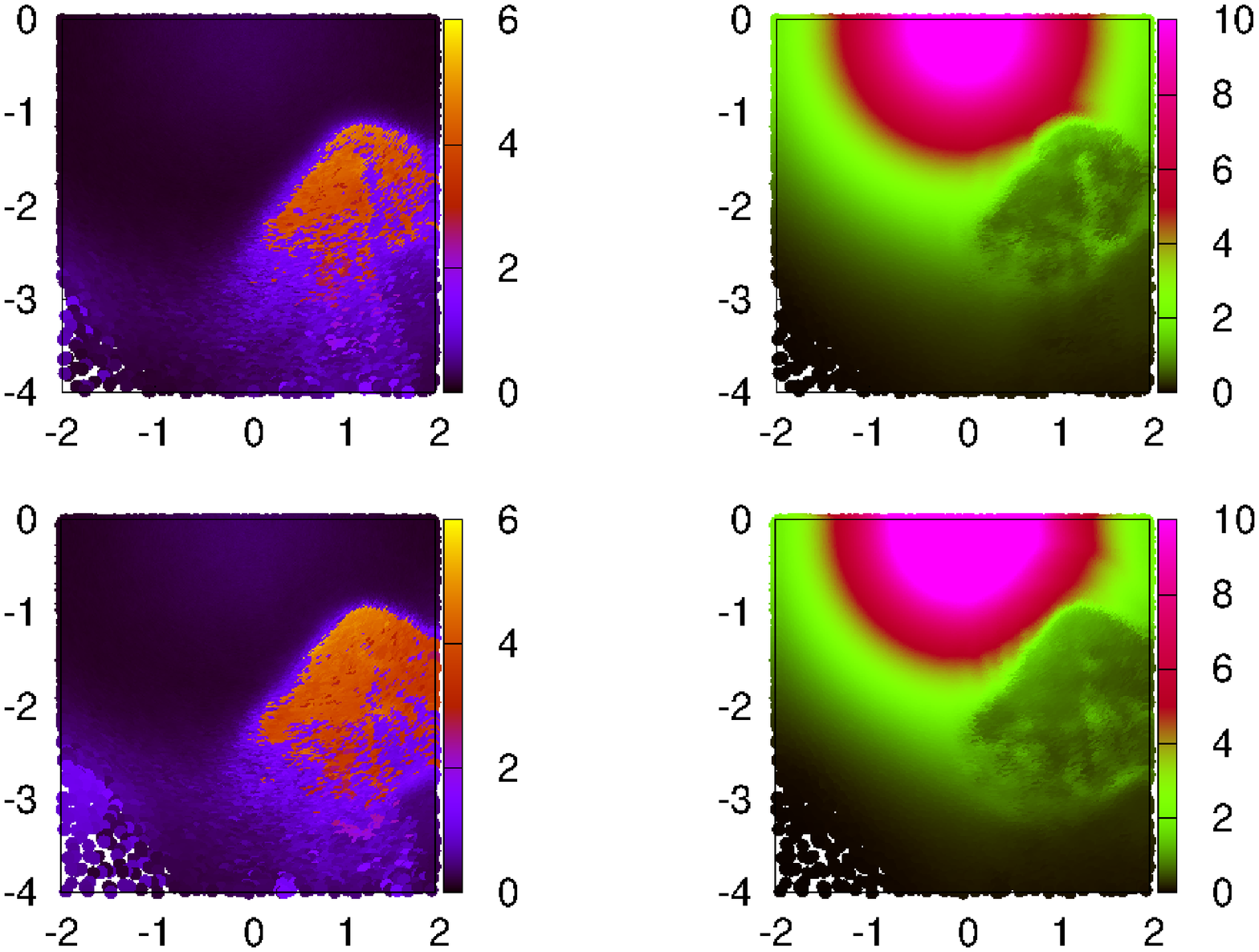}
\caption{Same as Figure~\ref{fig7} but for model C at times $t=4.28205$~s and $t=4.40775$~s with the nuclear reactions turned-on.  
}
\label{fig11}
\end{figure*}

Model C has the same initial angular velocity and kinetic rotational energy as model B but with $\Omega=\Omega_x$~going in an orthogonal direction to the initial displacement of the bubble. The overall view of the evolution of model C is shown in the third row of Figures~\ref{fig2} and \ref{fig4}. We clearly see that the mushroom-like structures are also rotating as they are dragged by the substrate. Actually they would look similar to that of models A and B in the comoving rotation frame. The impact of rotation becomes less acute at late times, because the expansion of the WD forces $\Omega$~to drop in order to preserve angular momentum, and by $t\ge 1.5$~s any angular movement has practically ceased while the radial expansion is strong. The last two panels of Figure~\ref{fig4} depict the jet formation for this case. We see that one of the effects of the rotation is to blur the component of the jet moving outwards, whose direction is now less defined. Interestingly, the chunk of the jet moving inwards appears to move eccentrically with respect the line defined by the head of the jet and the center of the remnant. As commented below this behavior is confirmed by a close-up inspection around the jet region. Despite of delivering the same amount of nuclear energy than in cases A and B model C shows some peculiarities at the time of the collision of the ashes. According to Figure~\ref{fig5}, the density of the fuel particle with maximum temperature at the jet region is lower than in the non-rotating case. At the fiducial temperature of $T_{max}=3\times 10^9$~K there is a $35\%$~reduction in density (column $12$~in Table~\ref{table1}) and the elapsed time for achieving $T_{max}$~is a bit shorter. As the conditions to form a detonation by the induction mechanism strongly depend on the density of the fuel \citep{nie97,dur06}, such reduction in density may have consequences to the further evolution of the WD. 

Model D was calculated assuming that $\Omega_x(r)$ follows the profile given by Equation~\ref{omega_1}. In this case, the core is rotating as a  rigid body with $\Omega_0=0.6$~s$^{-1}$ until a radius of $R_c=500$~km. Beyond $R_c$ a linear decay of $\Omega_x$ was assumed until it becomes negligible at the surface of the WD. These values of $\Omega_0$ and $R_c$ were chosen so that the rotational energy of model D at t=0~s is almost the same as that of models B and C. The evolution of model D, summarized in Figure~\ref{fig6}, was not very different than that of model C, but in this case the released nuclear energy during the deflagration was a slightly lower. As a consequence, the value of density in both, the center of the WD and the head of the jet, when the temperature $T_9=3$ is achieved for first time in the colliding region, are larger than in model C (see Table~\ref{table1}). This result supports the common idea that the total amount of released nuclear energy is the primary parameter that controls the thermodynamical conditions in the colliding region, which in turn determines the chances for the detonation of the core. Nevertheless, rotation also plays a role because, as shown below, the symmetry of the convergence of the ashes is also broken in model D. As in model C, the jet is digging tangentially into the core, thus changing the environmental conditions at the tip of jet.       

Finally, Figure~\ref{fig7} (models A and B) and Figure~\ref{fig8} (models C and D) show a close-up view of the jet region at the time when a detonation could have emerged if nuclear reactions were switched-on. In all cases we can see the jet traveling towards the core of the remnant. In  Figure~\ref{fig7} we see that the jet is broad ($\ge 1000~km$) due to the low resolution in this diluted region. The jet is accreting burnt material at its sides through an steady accretion shock and the pressure inside the jet is roughly balanced by the ram pressure of the infalling material. The detailed structure of the jet has been described in \cite{mea09} and \cite{sei09}, using a high-resolution axisymmetric approach. In models A and B the velocity field inside the jet is pointing to the center of the remnant. In these models, the jet is digging into the core (see the rightmost panels in Figure~\ref{fig7}), whereas at the tip of the jet there is a compressed region with higher temperature and density, and is made of fuel. According to previous studies it is just in this region  where there is some chance for a detonation \citep{sei09}~but knowing it unambiguously would require a 3D calculation with very high resolution ($\le 500$ m) which incorporates many pieces of physics (rotation among them). Models C and D rotating around a tilted axis show, however, a different behavior. According to Figure~\ref{fig8}, a jet is also born in the region of the collision of the ashes, but now the focusing effect is weaker  and, more important, the velocity field inside the jet is not pointing to the center of the core. This effect can also be seen in the color map of density where the jet is "bitting" the core laterally, which changes the density conditions found by precursor shock ahead the tip of the jet. Therefore, it seems that even a moderate amount of rotation may have an impact on the detonation formation and future studies should include this ingredient in the simulations.

\subsection{Core detonation ?}

The initiation of the detonation at the core edge, induced by the inwardly moving jet, has been analyzed with some detail by \cite{sei09} assuming that the jet is axisymmetric and restricting the calculation to a small region around the symmetry axis at the core interface. Their convergence study, with resolution between $\simeq 4$~km to $\simeq 0.1$~ km, showed that in some cases a detonation is formed via the gradient mechanism, when a given volume of fuel has a shallow enough density and temperature distribution so it burns in a time lower than the sound crossing time. A detonation occurs provided that the size of the burnt zone is larger than a critical value which strongly depends on density. According to \cite{rop07} that radius is  $\simeq 120$~km at $\rho= 3\times 10^6$~\dens~and $T_9= 2.3$, but it decreases to $\simeq 8$~km at $\rho= 10^7$~\dens~and $T_9=2.2$. In the calculations by \cite{sei09}, with a resolution of 0.5~km, these criteria are satisfied behind a compression front that moves transversely to the tip of the jet, facilitating the initiation of the detonation. 

Even though our resolution ($\simeq 50-70$~km, between the center of the core and the top of the jet) is not enough to elucidate if there is a detonation of the core, we have followed the penetration of the jet farther to compare the cases with and without rotation. Starting at some point at the plateau in Figure~\ref{fig3}, we resume the calculation {\sl switching-on} the nuclear reactions. In the same figure we see that the nuclear energy input becomes again relevant after t$=3.95$~s for model A and t$=3.97$~s for model C. In Figure~\ref{fig9} we show the color-map of temperature and density for model A at times $t_1=4.28032$~s and $t_2=4.41304$~s. It can be seen that the size of the high-temperature region grows in time as matter flows into the hot volume through the accretion shock. Nevertheless, the penetration of the apex of the jet in $\Delta t=t_2-t_1$~is hardly perceptible. On the top of the conical jet, but detached of it, there is weak shock-wave digging into the core. That wave is clearly visible in the density maps of Figure~\ref{fig9}, especially at $t_1=4.28032$~s. A one-dimensional cut around the symmetry axis showing the density and temperature profiles as a function of coordinate Z is shown in Figure~\ref{fig10}. These profiles are moving with phase velocity $\leq 1000$~km~s$^{-1}$, lower than the sound speed in the unburnt material, $C_s\simeq 4000$~km~s$^{-1}$. Thus, at our last computed time, $t_2=4.41304$~s, there was not a clear signal of detonation in model A.

In Figure~\ref{fig11} there are depicted the temperature and density color-maps of model C at times $t_1=4.28205$~s and $t_2=4.40775$~s.  As in Model A the nuclear reactions were turned-on above $t=2.5$~s and the color-map focusses on the region around the jet. The head of the jet is penetrating obliquely into the core while it grows in size owing to the nuclear combustion. As in model A a precursor shock is formed on top of the jet at time  $t_1$. This weak shock soon 
detaches from the subsonic jet and moves right into the core, distorting its spherical symmetry. Again there was not any clear indication of detonation in this region. The evolution of the released nuclear energy of models A and C at late times is shown in Figure~\ref{fig3} (thick red and blue lines respectively). As we can see both curves tends to saturate but the slope of the released nuclear energy is lower in model C. This is in agreement with the information shown in Figure~\ref{fig5} concerning the density of the fuel at a given temperature in the collision region. That density is lower in the rotating models and so they are the nuclear reaction rate and the released nuclear energy.   

If a detonation does not occur there are two possible outcomes for the remnant, 1) It simply remains oscillating while accreting a fraction of the ejected material made, for the most part, of unburnt particles. 2) the fall-back onto the core of some of the previously expulsed matter produces and accretion-shock. As the infalling material crosses the accretion-shock it is compressed and heated, leading to a second chance for a detonation of the core. This is the PRD explosion mechanism postulated by \citet{bra06}. We have tracked the evolution of models A and C for a while, until $t\simeq 12$~s. In both cases the evolution is quite complex showing several episodes of ashes collision while the core is oscillating. The follow-up of this long phase, probably leading to the PRD ignition of the core, is left for a future work.

\section{Discussion and Conclusions}

Multidimensional simulations of the thermonuclear explosion of white dwarfs have, for the most part, assumed that the event takes place in a non-rotating progenitor. This is actually a wrong hypothesis because the same accretion process which builds up the mass of the WD also transfers angular momentum from the disc to the compact object. As a result the WD could be spinning at angular velocities as high as $\Omega\simeq 2-3$~s$^{-1}$ at the moment of the ignition \citep{pie03,yoo05,dom06}. The reduction in the effective gravity caused by the centrifugal force changes the structure of the progenitor prior the explosion. The density profile is not longer spherically symmetric and, for the same central density, rotating white dwarfs could be  more massive than their non-rotating counterpart. If the amount of rotation is low,  to handle the explosion without the intervention of rotation is usually a good approximation. But in some particular scenarios even a slow rotation may have consequences in the development of the explosion. This is the case of the GCD scenario which relies on the efficiency of the convergence of the nuclear flame at the antipodes of the ignition point. A good efficiency is achieved provided the deflagration remains axisymmetric with respect to the line defined by the center of mass of the WD and the ignition point. Nonetheless, even a small amount of rotation may break the symmetry of the process because the Coriolis force is by nature non-axisymmetric and, being proportional to the velocity of the hot rising blobs, it can be strong. The Coriolis force is also acting in deflagrations arising from a multipoint ignition but in that case the stochasticity of the ignition may render its final impact less relevant.  

In this work we have studied the impact -via the Coriolis force- of having a moderate amount of rotation in the fate of the deflagration arising from a point-like ignition. Considering angular velocities $\Omega\simeq 0.4$~s$^{-1}$ is not at odds with the hypothesis of spherical symmetry, and allows a meaningful comparison of the results between rotating and non-rotating models. Such angular velocity is small enough to not change the density profile and equilibrium properties of the progenitor, but sufficient to appreciably push the hot blobs of incinerated material as they rise at velocities approaching the local sound speed. A larger value of $\Omega$~will make the Coriolis force even stronger and, in this sense, our calculations are rather conservative. On another note, moderate spinning WDs could be no so rare because: 1) exploding WDs rotating with $\Omega\leq 1$~s$^{-1}$~are expected for compact systems with total masses close to the canonical Chandrasekhar-mass limit. The physical structure of these WD would be similar to that of their non-rotating counterparts \citep{pie03,dom06}, and 2) as commented in the introduction the central deflagration of a fast spinning WD leaves a large amount of unburnt matter, which is incompatible with observations \citep{pfa10a}.

The effect of rotation is more pronounced when the rotation axis is orthogonal to the ascending line of the bubble, reinforcing the idea that the Coriolis force is the main agent behind the differences between spinning and non spinning models. In this respect, it is worth to mention that the value of the angle between the rotation axis and the ascending line of the bubble is  unknown. In order to have a better insight of realistic values of this angle, detailed 3D calculations of the progenitor including rotation are needed. 

For a similar released nuclear energy, models with low-moderate rotation show two distinctive features: 1) At the same elapsed time the temperature and density after the convergence of the deflagration at the antipodes are lower in rotating models, 2) as a result of the collision of the ashes a conical jet is born which grows in size through an accretion shock. In non-rotating models the jet is axisymmetric with its apex heading right to the center of the core. In rotating models, however, the conical symmetry is not so perfect and the jet is not longer pointing to the center of the core. Both features above suggest that, along the total released nuclear energy, rotation is also an important parameter to take into account to understand the GCD route to SNe Ia. 

Several studies have stated that the self-consistent detonation of the core of a white dwarf by the GCD mechanism is not easy, probably requiring  a fine tuning of the physical conditions at the convergence region of the flame. In particular the released nuclear energy during the deflagration phase has to be low \citep{rop07}, the geometry of the flame at late times should remain as axisymmetric as possible to also produce an axisymmetric jet \citep{mea09} and finally, giving birth to a detonation requires a precise preconditioning of the plasma at the tip of the jet (a shallow gradients of $\rho, T$~ in a  large enough volume) \citep{sei09}. Our exploratory calculations using low-moderate rotators suggest that detonating the core of the WD by the GCD mechanism could be even more difficult in spinning models because the symmetry of the deflagration is broken by the Coriolis force. Therefore a deeper understanding of the rotational features of a massive WD at the verge of the explosion is needed.

\acknowledgements

This work has been supported by the MINECO Spanish projects AYA2013-42762-P (D.Garc\'\i a-Senz) and AYA2011-22460 (I. Dom\'\i nguez) and  by the Swiss Science Foundation, PASC, European Research Council (FP7) under ERC Avanced Grant Agreement No. 321263 - FISH. (R. Cabez\'on and F.K. Thielemann). The Basel group is a member of the COST Action New CompStar. 

\clearpage
\appendix
\section{Flame implementation.}

Nuclear flames were handled using a reaction-diffusion scheme. Therefore, the actual thickness of the flame was artificially enlarged to the spatial local resolution of the simulation, roughly the smoothing-length parameter $h$. The actual conductivity coefficient was conveniently re-scaled to obtain a prescribed velocity for the flame, which was in turn computed from a flame subgrid model. Those cases where the scaling factor is one correspond to real microscopic laminar flames. The evolution of the flame is basically controlled by the energy equation:

\begin{equation}
\frac{du}{dt}=\frac{1}{\rho^2}~P~\frac{d\rho}{dt}+\frac{1}{\rho}\nabla\cdot(\kappa\nabla T)+\dot S_{nuc}
\end{equation}

\noindent
where $\kappa$ is the thermal conductivity and the remaining symbols have their usual meaning. Now consider the following linear transformations:

\begin{equation}
\mathbf r\longrightarrow \mathbf r'=a~\mathbf r;\qquad\qquad t\longrightarrow  t'=b~t, 
\end{equation}

\noindent
which maps the coordinate $\{r, t\}$ framework of the microscopic thermal wave to the laboratory $\{r',t'\}$ framework of the hydrocode.
Equation A1 can be written as:

\begin{equation}
du=\frac{1}{\rho^2}~P~d\rho+a^2~\frac{1}{\rho}\nabla'\cdot(\kappa~\nabla' T)(\frac{dt'}{b})+\dot S_{nuc}(\frac{dt'}{b})
\end{equation}

On the other hand, for a steady conductive flame moving with velocity $\mathbf v_c$:

\begin{equation}
\mathbf v_c\cdot \mathbf\nabla=\frac{\partial}{\partial t}
\end{equation}

Using the scaling transformations A2 in equation A4 gives:

\begin{equation}
a~\mathbf~v_c\cdot \mathbf\nabla'=b~\frac{\partial}{\partial t'}
\end{equation}

Therefore, only for $a=b$ the scaled flame will move just at the conductive velocity $v_c$. Thus, setting the ratio $a/b$ to:

\begin{equation}
\frac{a}{b}=\frac{v_f}{v_c}
\end{equation}

\noindent
leads to a prescribed effective velocity of the flame $v_f$. The practical procedure to implement the flame in the numerical scheme is:

\begin{enumerate}
 \item Make an estimation of parameter $a=(\frac{h}{\delta})$, where h is the smoothing length an $\delta$ is the actual microscopic width of the flame \citep{tim92}.
 \item Calculate the effective flame velocity $v_f$~using a subgrid model (see below). The parameter $b$ is then set to $b=a~(\frac{v_c}{v_f})$, where $v_c$ is the laminar flame velocity.
 \item Compute the conductivity $\kappa (\rho, T, X_i)$~and released nuclear energy rate and evolve the model using equation A3 as energy equation to describe the evolution of the internal energy and temperature.
\end{enumerate}

Because the scheme uses both the physical conductivity and nuclear reaction rates calculated through a 14-isotope network, it will provide the actual conductive velocity in the limit of very large resolution characterized by $a=b\simeq 1$. 

The nuclear network is able to follow all stages of the combustion, including the nuclear statistical equilibrium regime. Nevertheless, in order to save computing time the material is transformed isochorically to NSE elements once the temperature has raised over 3.5 billion degrees and density is higher than $5\times 10^7$~\dens. However, once the NSE state is achieved its evolution is followed again using the 14-isotope network. Such procedure allows to adequately follow the freeze-out of nuclear reactions.

\begin{figure}
\includegraphics[angle=-90,width=\columnwidth]{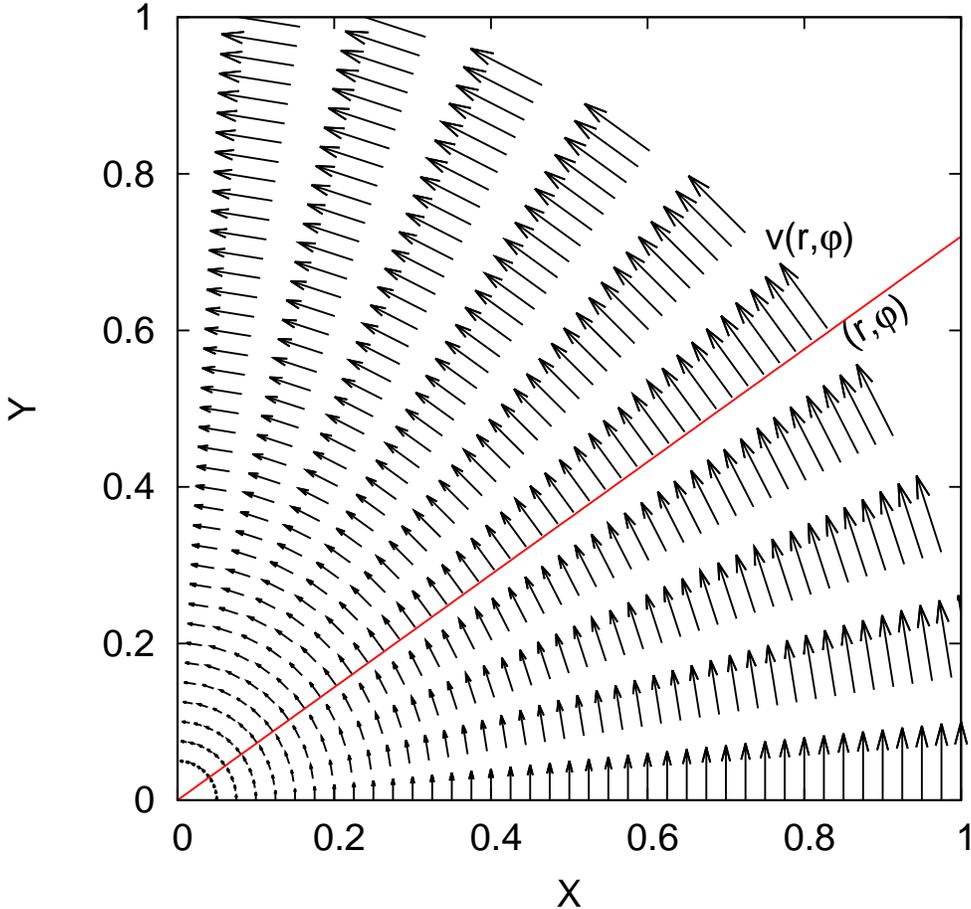}
\caption{Sketch of the toy-model used to estimate the characteristic turn-over particle velocity. The figure shows the velocity field calculated with Equation~\ref{velocityfield} and $n=1$~ as seen in the comoving frame of a particle located at the coordinate origin.}   
\label{figappendix1}
\end{figure}

A subgrid model was built by simply assuming that the size of the velocity fluctuation at the minimum scale-length solvable by the hydrocode is just the effective burning velocity at such scale \citet{dam40}. In order to evaluate the characteristic turn-over velocity of eddies with size $\simeq h$ we assume an isotropic velocity field around the particle. In cylindric coordinates the velocity at a distance $r$ of a given particle is described by:

\begin{equation}
\mathbf v(r,\varphi)=K r^n{\mathbf \tau}
\label{velocityfield}
\end{equation}

\noindent
where $\tau$~is the azimuthal unit vector. For $n=1$ expression A7 reduces to rigid rotation (see Figure~\ref{figappendix1}). Cases with $n=1/2$ and $n=1/3$ are representative of RT instability and turbulence \citep{nie97}.

Taking the curl of the above equation in cylindric coordinates and evaluating it at $r=h$ gives:

\begin{equation}
\mathbf\nabla\times v=(n+1)~K~h^{n-1}\mathbf k
\end{equation}

\noindent
where $\mathbf k$ is the unit vector along Z-axis. Constant $K$ in equation A7 becomes:

\begin{equation}
K= \frac{1}{n+1}~h^{1-n}\vert\nabla\times~v\vert_{SPH}
\end{equation}

\noindent
where $\vert\nabla\times~v\vert_{SPH}$ stands now for the curl calculated with the hydrocode. Therefore a rough estimation of the turn-over velocity at scale h is

\begin{equation}
v(h)\simeq K h^n=\frac{1}{n+1}~h~\vert\nabla\times~v\vert_{SPH}
\end{equation}

The turn-over velocity $v_t$ is finally taken $v_t=\alpha v(h)$, with $0< \alpha < 1$. In our calculations we use $n=1/3$, characteristic of turbulence and $\alpha=0.25$ fitted so that the total released nuclear energy during the failed explosion starting in a non-rotating single bubble is $\simeq 1.5\times 10^{50}$ ergs. At the moderate angular velocities considered in this work the estimated value of $v(h)$ is practically independent of $\Omega$.


\end{document}